\documentclass[12pt]{article}
\usepackage{amsmath}
\usepackage{amsfonts}
\usepackage{graphicx}
\usepackage{cite}
\usepackage{bbm}

\renewcommand{\thesection}
{\arabic{section} \hspace{-.5em}
}
\renewcommand{\thesubsection}
{\arabic{section}.\arabic{subsection}   \hspace{-.5em}
}
\renewcommand{\thesubsubsection}
{\arabic{section}.\arabic{subsection}.\arabic{subsubsection} \hspace {-.5em}
 }

\makeatletter
\renewcommand\section{
\@startsection{section}{3}{\z@}%
{-4.5ex\@plus -1ex \@minus -.2ex}%
{1.5ex \@plus .2ex}%
{\normalfont\large\bfseries\mathversion{bold}}}
\renewcommand\subsection{
\@startsection{subsection}{3}{\z@}%
{-3ex\@plus -1ex \@minus -.2ex}%
{0.7ex \@plus .2ex}%
{\normalfont\normalsize\bfseries\mathversion{bold}}}
\renewcommand\subsubsection{
\@startsection{subsubsection}{3}{\z@}%
{-3.25ex\@plus -1ex \@minus -.2ex}%
{1.5ex \@plus .2ex}%
{\normalfont\normalsize\itshape}}
\makeatother

\makeatletter \@addtoreset{equation}{section} \makeatother
\renewcommand{\theequation}{\arabic{section}.\arabic{equation}}

\makeatletter
\renewcommand{\appendix}{
\renewcommand{\thesection}{\Alph{section}  \hspace{-.5em}}
\renewcommand{\thesubsection}
{\Alph{section}.\arabic{subsection} \hspace{-.5em}}
\renewcommand{\thesubsubsection}
{\Alph{section}.\arabic{subsection}.\arabic{subsubsection} \hspace  {-.5em}}
\@addtoreset{equation}{subsection}
\renewcommand{\theequation}{\Alph{section}.\arabic{equation}}
\setcounter{section}{0}}
\makeatother

\renewcommand{\thefootnote}{\fnsymbol{footnote}}

\def\eqb         {  \begin{eqnarray}  }
\def\eqe           {  \end{eqnarray}  }
\def\nn               {  \nonumber  }
\def\bc       {\begin{center}}
\def\ec       {\end{center}}
\def\bra{\langle}
\def\ket{\rangle}
\def\comma      { \, , }
\def\period     { \, . }
\def\del        {  \partial  }
\def\delbar     {  \bar{\partial}  }
\def\half       {  {1\over 2}  }
\def\vecii#1#2     {{ #1 \choose #2 }  }
\def\matrixii#1#2#3#4            {\Bigl( \begin{array}{cc}#1&#2\\#3&#4
                                     \end{array} \Bigr) }
\def\vs#1   {\vspace*{#1 ex}}
\def\hs#1   {\hspace*{#1 ex}}                                     
\def\calN     { {\cal N} }   
\def\calA {{\cal A}}
\def\calM {{\cal M}}
\def\calO  {{\cal O}}
\def\ep{\epsilon}
\def\varep{\varepsilon}
\def\sql    {\sqrt{\lambda}}
\def\zbar   {\bar{z}}

\newcommand{\dilog}{\operatorname{Li_2}}

\newcommand{\bbR}{{\mathbb R}}
\newcommand{\bbZ}{{\mathbb Z}}                                  
\newcommand{\alg}[1]{{#1}}

\newcommand{\Eqn}[1]{& #1&}

\begin{document}
\def\papertitlepage{\baselineskip 3.5ex \thispagestyle{empty}}
\def\preprinumber#1#2#3{\hfill \begin{minipage}{2.6cm} #1
                \par\noindent #2
              \par\noindent #3
             \end{minipage}}
\renewcommand{\thefootnote}{\fnsymbol{footnote}}
\newcounter{aff}
\renewcommand{\theaff}{\fnsymbol{aff}}
\newcommand{\affiliation}[1]{
\setcounter{aff}{#1} $\rule{0em}{1.2ex}^\theaff\hspace{-.4em}$}
%
%
\papertitlepage
\setcounter{page}{0}
\preprinumber{}{UTHEP-653}{}
\baselineskip 0.8cm
\vspace*{2.5cm}
\begin{center}
{\Large\bf Gauge/string duality  \vspace*{0.5ex} \\
and thermodynamic Bethe ansatz equations}
\end{center}
\vskip 4ex
\baselineskip 0.7cm
\begin{center}
       
         Yuji  ~Satoh\footnote[3]{\tt ysatoh@het.ph.tsukuba.ac.jp}

\vskip 2ex
 
    {\it Institute of Physics, University of Tsukuba} \\
    {\it Tsukuba, Ibaraki 305-8571, Japan}
\end{center}
\vskip 7ex
%
\baselineskip=3.5ex

\begin{center} {\bf Abstract} \end{center}

\par\medskip
\ 
We review recent developments in the study of gluon scattering amplitudes
of the four-dimensional maximally supersymmetric Yang-Mills 
theory at strong coupling based on the gauge/string duality and its underlying 
integrability. The scattering amplitudes are given by 
the area of  minimal surfaces in five-dimensional anti-de Sitter 
space with a null polygonal boundary.
These minimal surfaces are described by integral
equations of the form of the thermodynamic Bethe ansatz equations.
Generalizing the result regarding the six-point amplitudes,
we observe a general connection between the minimal surfaces
and the homogenous sine-Gordon model, which is a class of 
two-dimensional integrable models associated with 
certain coset conformal field theories.  
We also demonstrate that the identification of the underlying 
integrable models is useful for analyzing the 
amplitudes by explicitly deriving an expansion of  the six-point amplitudes
around a special kinematic point.

%
%
%
%
%

\vspace*{\fill}
\noindent
November 2010%
\footnote[0]{
Contribution to the proceedings of RIMS Workshop ``Developments in Quantum Integrable Systems'', June 14-16, 2010, Research Institute for Mathematical Sciences, Kyoto, Japan;
RIMS Kokyuroku Bessatsu B {\bf 28} (2011) 171-192. }

\newpage
\renewcommand{\thefootnote}{\arabic{footnote}}
\setcounter{footnote}{0}
\setcounter{section}{0}
\baselineskip = 3.3ex
\pagestyle{plain}
\section{Introduction}

\subsection{Gauge/string duality and AdS/CFT correspondence}

The gauge/string duality emerged as a consequence of a natural development 
of the study of string solitons such as black holes (p-branes) and D-branes, 
and has been a central subject in string theory since mid-nineties. 
The studies of the matrix models for non-perturbative strings
and the quantum theory of black holes in string theory are notable
examples based on this duality. 
A basic picture of the duality is that at weak coupling the string 
solitons are described by open strings/gauge theories in flat space,
whereas at strong coupling they are described by closed strings/gravity.

In this talk, we focus on a particular form of the duality called
the AdS/CFT correspondence. This is the duality between
the string theory on five-dimensional
anti-de Sitter  space ($AdS_{5}$) times five-dimensional sphere ($S^5$)
and the four-dimensional $SU(N_c)$ super 
Yang-Mills (SYM) theory  which has the maximal $\calN =4$ supersymmetry.
The $\calN=4$ SYM theory is known to be a conformal field theory (CFT),
leading to the name of the correspondence. More precisely, the duality
states that the two theories are two facets of one entity: 
for $N_{c} \gg 1$, 
when the 't Hooft coupling $\lambda = g_{YM}^2 N_c $ is kept small, 
the theory is well described  by $\calN=4$ SYM, whereas
the description by the classical strings/gravity on $AdS_5 \times S^5$
is appropriate for $\lambda \gg 1$. 
On the string side, $\lambda  =  4 \pi g_{s} N_{c}= R^{4}/\alpha'^{2}$, 
where  $g_{s}$ is the string coupling, $R$ is the radius of 
$AdS_{5} \times S^{5}$ and $\alpha'$ is the inverse string tension.
The duality has been studied intensively for large $N_{c}$, but 
is expected to hold also for finite $N_{c}$.  
Schematically, 

\vs{1}    
\begin{center}
  \fbox{\parbox{5.2cm}{ \ String theory on $AdS_{5} 
  \times S^{5}$ 
   \vs{0.6}   \\ 
    \hs{6.7} $ \lambda = R^{4}/ \alpha'^{2}  \gg 1$   }} 
       \hspace*{1.5ex}
          \raisebox{0ex}{\parbox{2.4cm}{ \hs{3.3} dual 
          \\ \hs{3.5} $\Longleftrightarrow$  \\
             strong/weak}}
          \hspace*{0ex}  \fbox{\parbox{5.3cm}{ \ 4 dim. $\calN = 4$ $SU(N_{c})$ 
          SYM 
            \vs{0.6}  \\  \hs{7} $\lambda= g_{YM}^{2} N_{c} \ll 1$
          } }  
\end{center}
\vs{1} 
\noindent
This AdS/CFT correspondence 
has attracted much attention. 
First, the correspondence
embodies interesting long-standing theoretical ideas:
the equivalence between large $N_c$ gauge theory and string theory,
and the holography which states that quantum gravity is described
by lower dimensional non-gravitational theory.
Second, because of the strong/weak nature, one can study
the gauge theory at strong coupling by classical strings/gravity.
In fact, there are many works on the applications 
of the correspondence, for example, to low energy
hadron physics (holographic QCD and AdS/QCD), quark gluon plasma
and quantum entanglement. In particular, the application to gluon 
scattering amplitudes of $\calN=4$ SYM is the subject of this talk.
 
\subsection{Integrability underlying AdS/CFT correspondence} 

Among the works on the AdS/CFT correspondence,
the discovery of the underlying integrability 
in the planar limit ($N_{c} \gg 1$) opened up new dimensions.
Here, the integrability means on the string side
that the string sigma model on $AdS_5 \times S^5$ 
classically admits a flat current with a spectral parameter which 
generates infinitely many conserved charges. On the gauge side, 
it  means that
the dilatation operators representing the anomalous dimension 
for lower loops are, in the planar limit,  identified  with Hamiltonians of 
integrable quantum spin chains.
This discovery of the integrability enabled one to compare in detail  
the gauge and the string side beyond (nearly) supersymmetric sectors 
which are protected from quantum corrections at strong coupling.
Furthermore, assuming that this integrability holds for arbitrary coupling,
one can expect to 
  
\par\medskip
  $\bullet$ \ solve the four-dimensional SYM theory exactly including 
     the spectrum,
    
\par\smallskip
   $\bullet$ \ solve the important string theory on $AdS_5 \times S^5$, in spite that
   solving string  
   \\
   \hspace*{5.4ex} theory on curved space-{\it time} is generally very difficult,
      
\par\smallskip
   $\bullet$ \ prove (or disprove) the AdS/CFT correspondence,
  
\par\smallskip
   $\bullet$ \ deeply understand the AdS/CFT correspondence, and gain 
    useful insights into 
    \\
\hspace*{5.6ex}
    and, if necessary,  firm theoretical grounds for applications.


\par\medskip
As a state of the art of the study of the AdS/CFT correspondence based 
on the integrability,
there is now a proposal: 
 { \it the spectrum of the string theory on
$AdS_5 \times S^5$ and the four-dimensional $\calN=4$ $SU(N_c)$ 
SYM theory for large $N_c$ and arbitrary coupling $\lambda$ is obtained 
by solving a certain set of equations.}
(For details, see the article by Prof. Tateo \cite{Tateo-san}.)
This set of equations takes the form of the
thermodynamic Bethe ansatz (TBA) equations or the Y-system, 
which appear in the study of finite-size effects of (1+1)-dimensional 
integrable models. This proposal has been checked up to 4-loop order
for a simple single-trace operator called the Konishi operator.
The spectrum of this operator at 5-loop order has also been computed by
using the L\"{u}scher formula.

Given this impressive progress  
in understanding  the AdS/CFT correspondence, one may also expect 
that the integrability must shed new light on applications of the
correspondence. It turned out that this is indeed the case: 
Based on their earlier work \cite{Alday:2007hr} that 
gluon scattering amplitudes
of $\calN=4$ SYM at strong coupling  for large $N_{c}$ are given 
by minimal surfaces in $AdS_5$, 
Alday and Maldacena
initiated a program to compute the  amplitudes
by using the integrability \cite{Alday:2009yn}.
In this program, 
the minimal surfaces in $AdS_5$ are described by
a set of integral/functional equations. Surprisingly, these again take
 the form of the TBA equations/Y-system 
 \cite{Alday:2009dv,Alday:2010vh,Hatsuda:2010cc}. However, 
 the TBA equations/Y-system here 
are different from those for the spectral problem mentioned above.
Thus, the gluon scattering amplitudes at strong coupling/minimal surfaces 
in $AdS_{5}$ provide
another example in which one finds unexpected connections
between the AdS/CFT correspondence and the TBA equations/Y-systems. 
Schematically, 
\vs{1}
\bc
\fbox{\parbox{9.2cm}{ \ Gluon Scattering Amplitudes at Strong Coupling \ }}

\vs{0.5}  \hs{10}
$\Uparrow$  \quad ref. \cite{Alday:2007hr} 

\vs{0.5}
\fbox{\parbox{5cm}{ \ Minimal Surfaces in $AdS_5 $  \ }}

\vs{0.5} \hs{13}
 $\Uparrow$  \quad refs. \cite{Alday:2009yn,Alday:2009dv,
 Alday:2010vh,Hatsuda:2010cc} 
 
\vs{0.5}
{\fbox{\parbox{7.8cm}{  \ Thermodynamic Bethe Ansatz Equations \ }} }

\ec

\vs{0}
\subsection{Plan of talk}

 In this talk, we next give a brief summary on the scattering amplitudes of  
 $\calN=4$ SYM both at weak and strong coupling in section 2.
 We then review developments in 
 the study of the scattering amplitudes based on
 the AdS/CFT correspondence and its underlying integrability in section 3.
 (See \cite{Alday:2009yn,Alday:2009dv,Alday:2010vh,Hatsuda:2010cc,
 Hatsuda:2010vr,Alday:2010ku} and  references therein.)
 We  move on to a discussion on the integrable models
 and the CFTs associated with the TBA equations/Y-systems
 for the minimal surfaces in section 4. In particular, we observe 
 \cite {Hatsuda:2010cc} 
 that the TBA equations 
 for the minimal surfaces in $AdS_{3}$ and $AdS_{4}$, corresponding to some
 kinematic configurations, coincide with 
 those of the homogeneous sine-Gordon (HSG) model  
 \cite{FernandezPousa:1996hi},
 which is a class of  (1+1)-dimensional integrable models associated with 
certain coset or generalized parafermion CFTs.  
This generalizes the connection \cite{Alday:2009dv} between the minimal 
surfaces in $AdS_{5}$ for the six-point amplitudes and the  
$\bbZ_{4}$-symmetric integrable model.
  Finally, we derive  
 an expansion of the six-point amplitudes 
near the CFT limit corresponding to a special kinematic
point \cite{Hatsuda:2010vr} in section 5.
  This demonstrates that the identification of the associated
 integrable models and CFTs is actually useful for analyzing 
 the amplitudes at strong coupling.
We conclude with a summary and discussion on future directions
 in section 6.

\section{Gluon scattering amplitudes of $\calN=4$ SYM}

\subsection{Amplitudes at weak coupling  and BDS conjecture}
Let us begin with a brief summary of the gluon scattering amplitudes
of  4-dimensional $\calN=4$ SYM theory at weak coupling
$\lambda = g_{YM}^{2} N_{c} \ll 1$ . For a review regarding
section 2, see for example \cite{Alday:2008yw}.
This theory contains a gauge field $A_{\mu}$ $(\mu=0,...,3)$, 
six scalars $\Phi^{i}$ $(i=1, ...,6)$
and four fermions $\psi^{a}$ $(a=1,...,4)$. All the fields take values 
in the adjoint representation of $SU(N_{c})$. This theory is obtained
by dimensional reduction from  10-dimensional $\calN=1$ SYM theory.
The theory also has the superconformal symmetry
$\alg{psu}(2,2 \, | \, 4)$. The bosonic part $\alg{su}(2,2) \oplus \alg{su}(4) 
\simeq \alg{so}(2,4) \oplus \alg{so}(6)$ represents the 4-dimensional
conformal symmetry and the R-symmetry. Note that $SO(2,4)$
and  $SO(6)$ are the isometries of $AdS_{5}$ and $S^{5}$, respectively.

In the planar limit $N_{c} \to \infty$ with the 
  't Hooft coupling $\lambda $ kept  small,   
an interesting conjecture is known 
that  the maximally helicity violating 
(MHV)  amplitude has a simple iterative structure to all orders in perturbation.
This is called the BDS (Bern-Dixon-Smirnov)  conjecture. To state the content
of the conjecture, we first note that the $n$-point  amplitudes at $L$-loop order
 are decomposed 
as follows:
\eqb
 A^{(L)}_n = N^{(L)} \sum  \mbox{\rm (color factor)}  \times \calA^{(L)}_n
 + \mbox{\rm (multi-trace part)} \period \nn
\eqe
The remainder $\calA^{(L)}_{n}$ after the color factor is factorized is called 
the color-ordered amplitudes. In the planar limit, the multi-trace part is neglected.
From the color-ordered amplitudes, the tree amplitudes are further
factorized,
\eqb
  \calA^{(L)}_n = \calA_n^{\rm tree} \times \calM_n^{(L)} \period \nn 
\eqe
The BDS conjecture states that the scalar part at $L$-loop order
$\calM_n^{(L)}$ is given by an iteration of the 1-loop result
through the generating function,
\eqb
  \calM_n 
= \exp\left[  \sum_{k=1}^\infty a^k f^{(k)}(\ep) \calM_n^{(1)}(k\ep)
 + C^{(k)} + \calO(\ep) \right] \comma \nn 
\eqe
where $a= {\lambda} (4\pi e^{-\gamma_E})^\ep/{8\pi^2}$ with $\gamma_E$ being
Euler's constant is the coupling constant customarily used
in loop calculations, and $f^{(k)}(\ep)$ and $C^{(k)}$ are certain constants 
independent of  external momenta. 
Note that $\calN=4$ SYM
is a massless gauge theory and thus one has to regularize the infrared 
divergences of the amplitudes by an infrared cut-off and 
dimensional regularization with $d=4-2\ep$.
The divergences are canceled 
in infrared safe quantities, which are obtained by combining the
scattering amplitudes. This conjecture has been checked
up to higher loops for 4- and 5-point amplitudes.

Probably, it is illuminating to see a concrete example of the 4-point
amplitudes,
\eqb\label{BDS4}
  \calM_4 = \calM_4^{\rm div} \times \exp\biggl[ \,  \frac{1}{8} f(\lambda) \bigl(\ln \frac{s}{t} 
 \bigr)^2
 + {\rm const.} \, \biggr] \comma 
\eqe
where $\calM_4^{\rm div}$ is the divergent part and  $s,t$ are the Mandelstam variables. A remarkable fact is that all the coupling dependence is encoded in 
the cusp anomalous dimension, 
\eqb
    f(\lambda)  = \frac{\lambda}{2\pi^2} (1- \frac{\lambda}{48} + \cdots)
    \period \nn
\eqe    

\subsection{Amplitudes at strong  coupling from AdS/CFT correspondence}

Now, let us move on to a discussion in the strong coupling region with 
$\lambda \gg 1$. Based on the AdS/CFT correspondence,
Alday and Maldacena argued that the scalar part of the amplitudes
at strong coupling is obtained by evaluating 
the action of the 
string sigma model on $AdS_5$ for certain classical string solutions 
\cite{Alday:2007hr}.
The saddle-point action gives the area of 
 minimal surfaces, meaning that the amplitudes
are given by the area of the minimal surfaces
in $AdS_5$:
\eqb
  \calM_n  \ \sim  e^{-S}  \ = \  e^{-\frac{\sql}{2 \pi} {\rm (Area)}}
  \period \nn
\eqe
The momentum dependence of the amplitudes come from 
the boundary condition of the minimal surfaces. This is analyzed 
by making use of T-dual transformations, and it turns
out that the surfaces have to end on a polygonal boundary 
on the boundary of $AdS_5$. See fig. 1.  There, each side of the 
polygon is null and corresponds to the momentum of an external
particle. The momentum conservation $\sum_i p^\mu_i =0$
implies that the boundary is closed. Thus, denoting the 
vertices of the polygon by $x^\mu_i$ (in terms of
the  Poincar{\'e} coordinates defined below), one has
\eqb\label{xp}
   \Delta_i x^\mu := x^\mu_i - x^\mu_{i+1} = p^\mu_i \period 
\eqe 
The $S^5$ part is expected to contribute to
 subleading terms of the amplitudes, but 
its role  is not so clear.

\begin{figure}[t]
   \bc
   \includegraphics*[width=5.7cm]{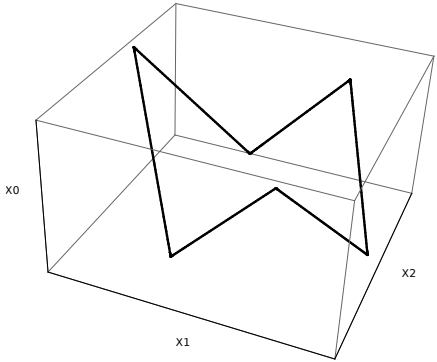}
   \caption{Boundary of a minimal surface on the AdS boundary.
    The axes in the figure represent the Poincar{\'e} coordinates
    $(x^1,x^2,x^0)$. }
   \ec
   \vs{-2}
\end{figure}

Again, it would be  illuminating to see a concrete example 
of the 4-cusp minimal surfaces describing the 4-point amplitudes. 
For this purpose, we first parametrize  $AdS_5$ as a hypersurface
in $\bbR^{2,4}$ defined by
\eqb\label{AdS5}
  \vec{Y}\cdot\vec{Y} :=
  -Y_{-1}^2-Y_0^2+Y_1^2+Y_2^2+Y_3^2+Y_4^2=-1 \period 
\eqe
The equations of motion for the string coordinates 
$\vec{Y}(z,\zbar)$
are
\eqb\label{eom}
 \del\delbar\vec{Y}
-(\del \vec{Y}\cdot\delbar\vec{Y})\vec{Y}=0 \comma 
\eqe
whereas the Virasoro constraints are
\eqb\label{Virasoro}
  (\del \vec{Y})^2= (\delbar \vec{Y})^2 =0  \period 
\eqe
Here, $z,\zbar$ are the world-sheet coordinates and
$\del = \del_z, \delbar=\del_{\zbar}$. 
One can check that these are the equations of the minimal surfaces.

A simple solution to (\ref{eom}) and (\ref{Virasoro}) is 
\eqb\label{4cuspsol}
 \matrixii{Y^{-1}+Y^4}{Y^1+Y^0}{Y^{1}-Y^0}{Y^{-1}-Y^4} 
= \frac{1}{\sqrt{2}} \matrixii{e^{\tau+\sigma}}{e^{\tau-\sigma}}{
-e^{-\tau+\sigma}}{e^{-\tau-\sigma}} \comma 
\eqe
with $Y^2=Y^3 =0$ and $z=\tau+ i\sigma$. 
To see what surface is described by this solution, let us introduce
the Poincar{\'e} coordinates defined by 
\eqb
  Y^\mu =: \frac{x^\mu}{r} \comma \quad
  Y^{-1}+Y^4 =: \frac{1}{r} \comma \quad
  Y^{-1}-Y^4 = \frac{r^2+x^\mu x_\mu}{r}
  \comma \nn
\eqe
where $\mu = 0,1,2,3$. In this coordinate system, the boundary 
of $AdS_5$ is located at  $r=0$.
Since the surface  is embedded in the $AdS_3$ subspace 
parametrized by $r$ and $ x^\pm := x^0 \pm x^1$, 
the external momenta given by (\ref{xp}) are in $\bbR^{1,1}$
and correspond to a restricted kinematic configuration.
Substituting the solution into these coordinates,
one can draw the picture of the surface as in the left figure in fig. 2.  
In the figure,
$AdS_3$ is represented as a  solid cylinder,  
where the radial direction
is parametrized by $r$ 
and the boundary of the cylinder by $x^{0}$ and $x^{1}$.
The $AdS$ boundary at infinity has been  mapped to
the boundary of the solid cylinder. To have a closer look, let us go around
the world-sheet $z$-plane far from the origin (middle figure in fig. 2). 
We then find that
the region far from the origin in the first quadrant is mapped to a
neighborhood of the origin in the $(x^+,x^-)$-plane (right figure in fig. 2).
Similarly, the regions far from the origin in the second, third, and fourth
quadrants are mapped to neighborhoods of 
$(x^+,x^-) = (\infty,0), (\infty,\infty),(0,\infty)$, respectively.
Thus, as we cross the real or the imaginary axis of the $z$-plane,
the boundary of the surface jumps from one cusp to another
and draws null lines.  In this way, the solution describes
a minimal surface with a polygonal boundary
consisting of four cusps and four null sides.
The solution (\ref{4cuspsol}) is thus the solution which 
we are looking for.

\begin{figure}[t]
   \hspace*{-2ex}
   \bc
   \includegraphics*[width=4.7cm]{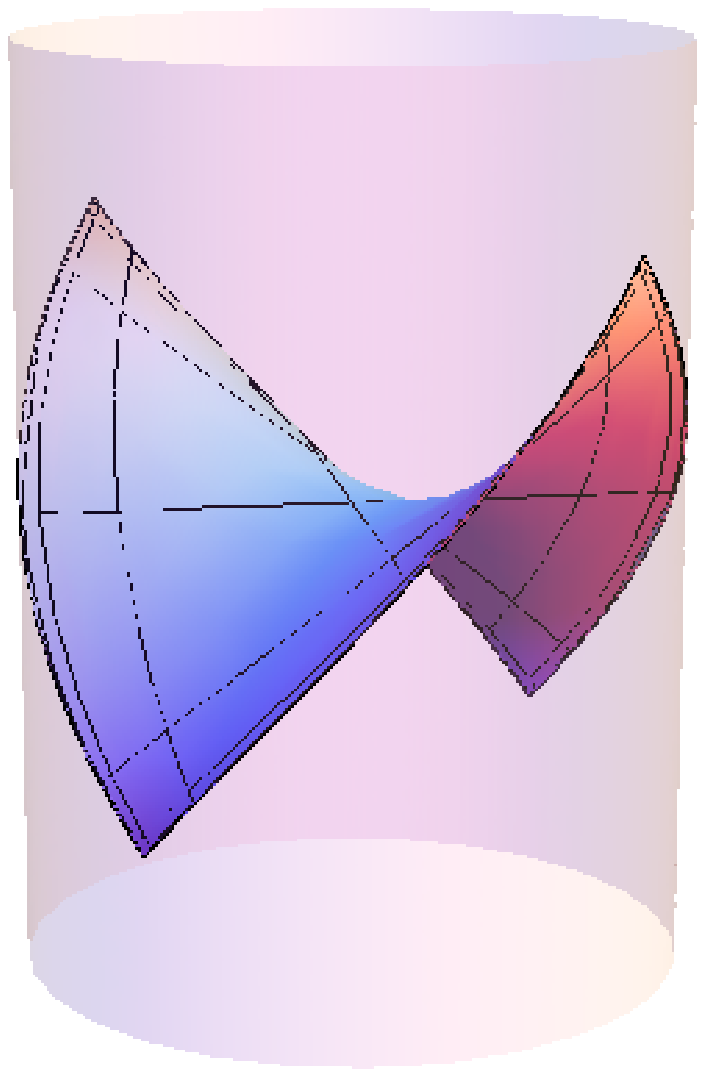}
   \hs{3.5} 
   \includegraphics*[width=3.8cm]{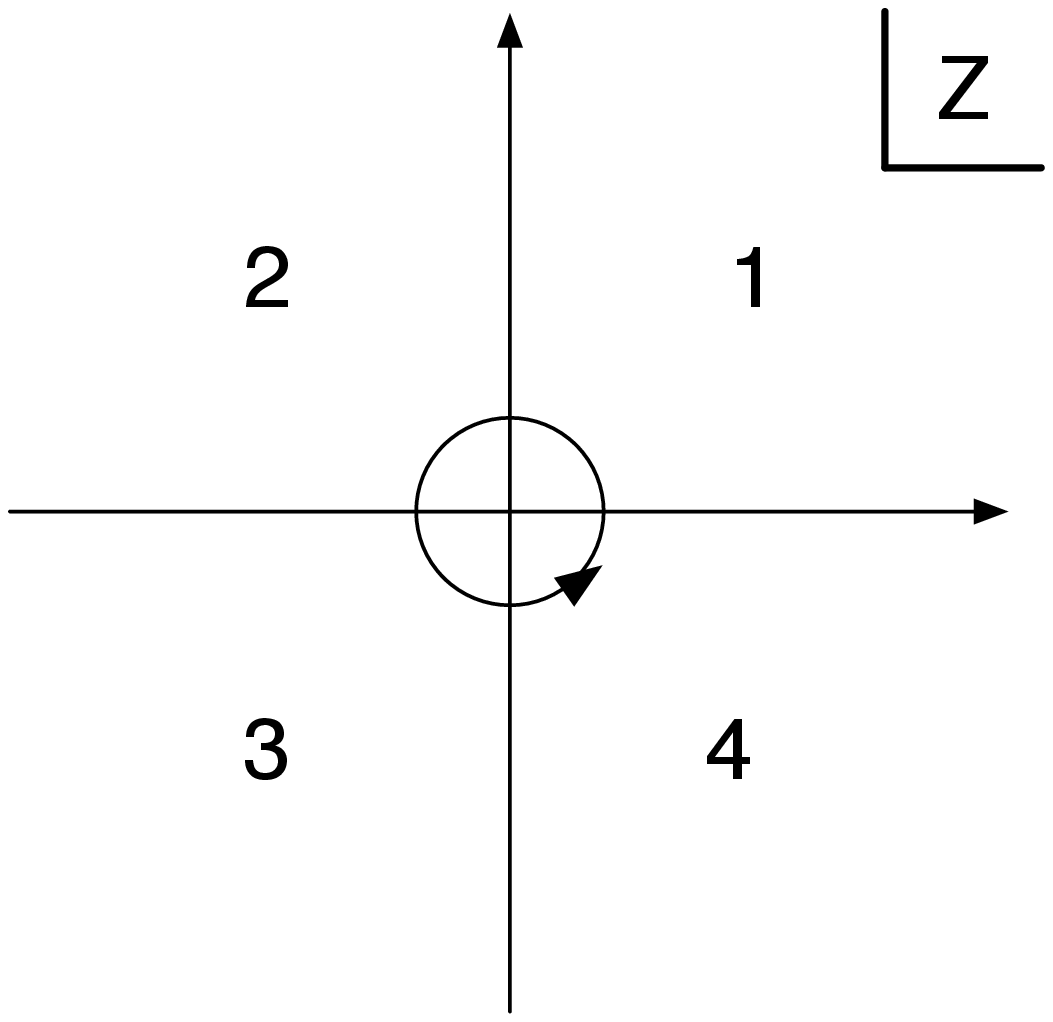}
   \hs{2.5}
   \includegraphics*[width=3.8cm]{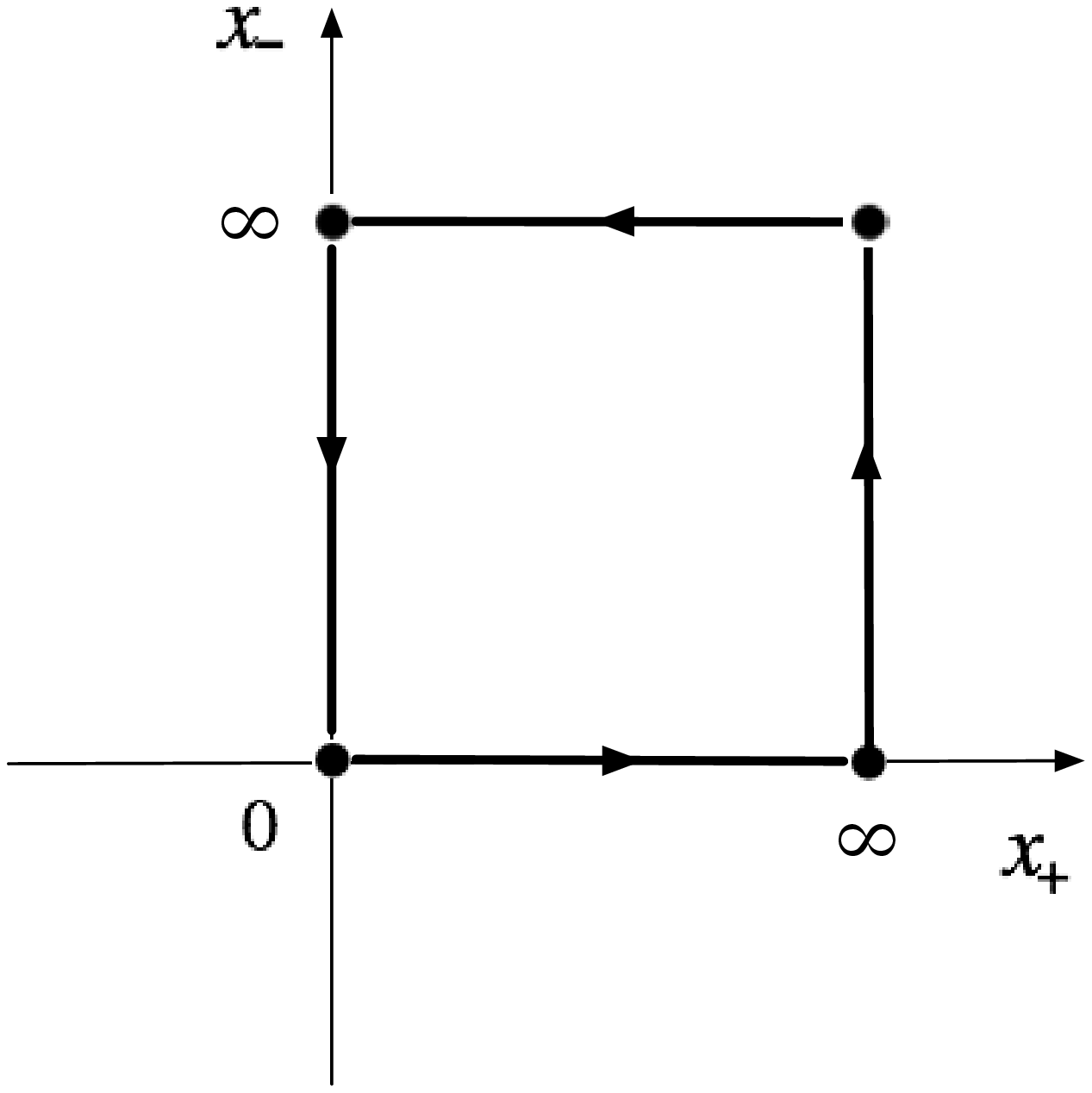}
   \caption{4-cusp minimal surface in $AdS_{3}$ (left), world-sheet $z$-plane
  (middle), and surface boundary in the $(x^+,x^{-})$-plane (right).}
  \ec
 \vs{-2}
\end{figure}

Minimal surfaces corresponding to more general
kinematic configurations are obtained by $SO(2,4)$
transformations. 
According to the prescription by Alday and Maldacena, 
the area of those minimal surfaces then gives the 4-point amplitudes
at strong coupling. Since the surfaces extend to infinity, their area
diverges and hence has to be regularized. Either by 
a dimensional regularization with $d= 4-2\ep$ or by 
a  cut-off regularization with $r > \ep'$, one finds that 
\eqb\
\calM_4  \ \sim 
   \calM_4^{\rm div} \times \exp\biggl[ \,  \frac{1}{8} f(\lambda) \bigl(\ln \frac{s}{t} 
 \bigr)^2
 + {\rm const.} \, \biggr] \comma \nn
\eqe
where  $ f(\lambda) = \sql/\pi$ and $ \calM_4^{\rm div}$ is the divergent term.
Remarkably, this has the same structure as the BDS formula (\ref{BDS4})
including the divergent term. Furthermore,
the value of $f(\lambda)$
here precisely agrees with the cusp anomalous dimension at strong
coupling which has been computed in the spectral problem of the AdS/CFT 
correspondence. 

\subsection{Insights from strong coupling computation}
This agreement  of the 4-point amplitudes is very impressive. 
Moreover, the study of the amplitudes at strong coupling provided
very useful insights into the weak coupling side and 
led to deeper understanding. 
First, it is known that  minimal surfaces in $AdS$ give expectation
values at strong coupling of Wilson loops  along the boundary of the surfaces.
Thus, the above discussion implies
that, at strong coupling, the amplitudes are the same  as the expectation 
values of the null polygonal Wilson loops.
A natural question here is whether  this is also the case on the weak 
coupling side. It then turned out that the answer is yes, as far as 
comparison is possible.
This correspondence between the amplitudes and the null polygonal
Wilson loops are now called the Amplitude/Wilson loop duality.

 Second, a detailed analysis on the strong coupling side
 for $n$-point amplitudes with $n\to\infty$ revealed  that the 
 BDS formula needs to be modified at strong coupling.
 Subsequent studies  confirmed
 that this is also the case  on the weak coupling side for $n \geq 6$. 
 Now, the deviation from the BDS formula is called
 the remainder function. Given the BDS formula,
 computing the amplitudes is equivalent to computing
 the remainder function. The remainder function is thus
 a central quantity in this subject.

 Third, 
 the computation on the strong coupling side
 manifests the conformal symmetry 
 in a sort of momentum space of the SYM theory, which is 
 (a part of) the T-dualized target space represented by  $x^{\mu}$. 
 This facilitated again the studies on the weak coupling
 side.  Together with earlier observations,
 the results support the existence of this symmetry also at weak coupling,
 which is now called the dual conformal symmetry.

 The dual conformal symmetry is natural on the strong coupling/string
 side, since it  corresponds to a T-dual symmetry or the Yangian symmetry
 of the string sigma model.  Moreover, once the existence of this symmetry
 is assumed, that leads to important consequences:
 The Ward identity associated with this symmetry
 strongly constrains the form of the amplitudes and,
 for the $n$-point amplitudes with $n \leq 5$, the BDS
 formula turns out to be unique. For $n \geq 6$, the
 Ward identity allows, in addition to the BDS form, functions
 of the cross-ratios of the cusp coordinates $x_i^\mu$, which
 are dual-conformal invariants and related to external momenta by (\ref{xp}). 
 The remainder function is thus a function of the cross-ratios.
 
\section{Minimal surfaces in AdS  and integrability}

In the following, we focus on the strong coupling/string side.
Triggered by the computation of the 4-point amplitudes in \cite{Alday:2009yn},
there were many attempts at constructing the minimal
surfaces with more than 4 cusps. For example, 
 cusp solutions are numerically studied in \cite{Dobashi:2008ia}, and
 a special 6-cusp solution is constructed by systematically 
 analyzing finite-gap solutions and their degenerate limits in \cite{SS}.
However, it turned out that  it is very difficult to construct the minimal surfaces
with the special null polygonal boundary.

Then, 
Alday and Maldacena initiated a program of general construction based on 
integrability \cite{Alday:2009yn}. 
They reduced the analysis of the minimal surfaces
to that of the Hitchin system and used  related results in the study 
of the wall-crossing phenomena of $\calN=2$ SYM. Roughly speaking,
they showed how to patch the 4-cusp solution (\ref{4cuspsol})
to form the general $n$-cusp solution. What is interesting 
is that the explicit form of the solution still is not available, but 
it is possible to compute the amplitudes.  In this way, 
they analyzed the 8-point amplitudes corresponding to
 the 8-cusp solution in $AdS_3$.

Subsequently, the 6-cusp solution in $AdS_5$ was discussed 
in \cite{Alday:2009dv} and, together with an argument on 
general cusp solutions, 
the 10- and 12-cusp solution in $AdS_3$ were discussed 
in \cite{Hatsuda:2010cc}. The 
general construction of the $n$-cusp solution in $AdS_5$
was then given in \cite{Alday:2010vh}. Below, we would like to 
explain this general construction. 
For simplicity, we focus on the case of $AdS_3$.

\subsection{General null polygonal solutions in $AdS_{3}$}
The first step in this construction is to reduce the analysis
of the classical solution of the $AdS$ sigma model to
that of the Hitchin system. This step is called the Pohlmeyer
reduction. Mathematically, this is equivalent to 
considering the evolution of a moving frame.
Concretely, one first takes a basis in $\bbR^{2,2} \supset AdS_3$, 
$q=(\vec{Y}, \del \vec{Y}, \delbar \vec{Y}, \vec{N})^t$, where
$N_a := \frac{1}{2} e^\alpha \epsilon_{abcd} Y^b \del Y^c \delbar Y^d$
and $ e^{2 \alpha} := \frac{1}{2} \del \vec{Y} \cdot \delbar \vec{Y}$.
$Y_a$ $(a=-1,0,1,2)$ are the embedding coordinates  which parametrize
$AdS_3$ similarly to  (\ref{AdS5}).
Since $q$ spans a frame at each point of $\bbR^{2,2}$, derivatives
of $q$ are again expressed by linear combinations of the elements of
$q$ itself. It is then possible to write the original equations of 
motion (\ref{eom}) and the Virasoro constraints (\ref{Virasoro})
in the form of an evolution equation $(d +U)q = q$,
where $d$ stands for the world-sheet derivative and $U$ is a certain matrix.

Furthermore,  decomposing  $SO(2,2)$ vectors by 
products of $\alg{su}(2)$ spinors through 
 $\alg{so}(4) \cong \alg{su}(2) \oplus \alg{su}(2)$ and
 introducing a complex parameter $\zeta$ (spectral parameter),
 the evolution equation is rewritten as
 \eqb\label{dpsi}
    0 =  \Bigl[ d + B(\zeta) \Bigr] \psi \comma  
 \eqe
where $\psi$ is a spinor related to $q$, 
\eqb
 B_z(\zeta) 
   = \matrixii{\half \del \alpha}{-\frac{1}{\zeta} e^{\alpha}}{-\frac{1}{\zeta}
    e^{-\alpha} p}{-\half \del \alpha} , \quad 
   B_{\zbar}(\zeta)
   = \matrixii{-\half \delbar \alpha}{-{\zeta} e^{-\alpha} \bar{p}}{-{\zeta}
    e^{\alpha}}{\half \delbar \alpha} \comma \nn
\eqe
and $p  := -2 \del^2 \vec{Y} \cdot \vec{N}$. It turns out that $p$ is 
holomorphic in $z$. We further decompose the connection $B(\zeta)$
according to the grading with respect to $\zeta$,
\eqb
   B_z(\zeta) = : A_z +  \frac{1}{\zeta}\Phi_z \comma
   \quad
   B_{\zbar}(\zeta) =:  A_{\zbar }+  \zeta \Phi_{\zbar}
   \period \nn
\eqe
The evolution equation of $q$ or (\ref{dpsi}) implies that
the original non-linear equations of the string sigma model
have been linearized.

The compatibility condition of (\ref{dpsi}),
$ 0 = [ \del + B_z, \delbar + B_{\zbar} ] $, is expressed as
\eqb\label{Hitchin}
  D_{\zbar} \Phi_z = D_z \Phi_{\zbar} = 0 \comma \quad
  F_{z\zbar} + [\Phi_z, \Phi_{\zbar}] =0 \comma 
\eqe
with $D \Phi = d \Phi + [A, \Phi] $. This is nothing but the
$\alg{su}(2)$ Hitchin system, which is obtained by dimensional
reduction of the 4-dimensional self-dual (instanton) equations.
In the $AdS_{5}$ case, one similarly finds the $\alg{su}(4)$ Hitchin system.
Tracing back the argument, a solution to the Hitchin system
gives a proper solution to (\ref{dpsi}), and it then  gives
$q$ and a solution to the original string equations $\vec{Y}$.
The formula to reconstruct $\vec{Y}$  is
\eqb
  Y_{a\dot{a}} := \matrixii{Y^{-1}+Y^2}{Y^1+Y^0}{Y^{1}-Y^0}{Y^{-1}-Y^2} 
    = \Psi(\zeta=1) M \Psi(\zeta=i) \comma \nn
\eqe
where $\Psi(\zeta)=(\psi_1,\psi_2)$ with $\psi_{1,2}(\zeta)$ being
properly normalized independent solutions  of (\ref{dpsi}), and
$M$ is a certain matrix.

Now, we are ready to discuss  general cusp solutions.
We recall that the number of the cusps
is even in $AdS_3$.  
For our purpose, we first make changes of variables for 
the world-sheet coordinates by $ dw = \sqrt{p(z)} dz$
and for the potential 
$\alpha$ by $\hat{\alpha} = \alpha - \frac{1}{4} \ln p\bar{p}$.
In term of $\hat{\alpha}$,  the compatibility condition of (\ref{dpsi}), 
or (\ref{Hitchin}), reduces to  the sinh-Gordon equation
$
 \del_w\del_{\bar{w}} \hat{\alpha} -2 \sinh \hat{\alpha} =0 \period 
$
Here, we note that the linear problem with $\hat{\alpha} = 0$ gives
the 4-cusp solution (\ref{4cuspsol}) (with $Y^4 \to Y^2$) in 
the $w$-plane. Thus, if we take $p(z)$ to be a polynomial of degree $n-2$, i.e., 
$p(z) = z^{n-2} + \cdots$,  and find a solution where $\hat{\alpha} \to 0$
as $|w| \to \infty$, that is 
the $2n$-cusp solution  in the original $z$-plane. 
The reason is as follows: First, since $w \sim z^{n/2}$ for large $|z|$,
if we go around the $z$-plane once far from the origin,  we
go around the $w$-plane $n/2$ times (fig. 3). The solution with
$\hat{\alpha} \to 0$ as $|w| \to \infty$ produces one cusp
in each quadrant, as explained below (\ref{4cuspsol}), and thus, 
from the point of view of the $z$-plane, the solution
has $4 \times n/2 = 2n$ cusps.
In a canonical form, the polynomial $p(z)$ has
$2(n-3)$ real parameters, which agrees with the number 
of independent cross-ratios in the $2n$-point scattering
for the kinematic configurations corresponding to
$AdS_3$.

\begin{figure}[t]
   \bc
   \includegraphics*[width=5.1cm]{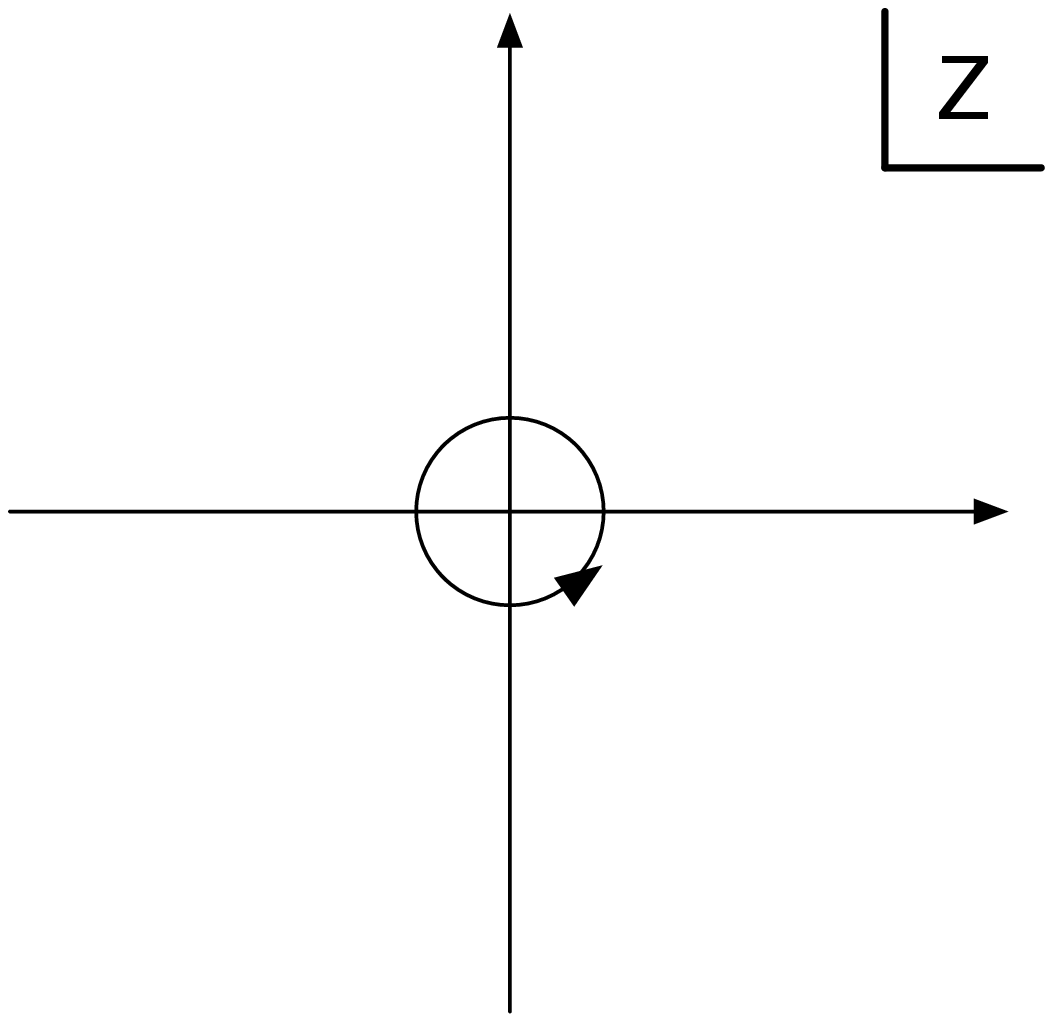}
   \hs{9}
   \includegraphics*[width=5.1cm]{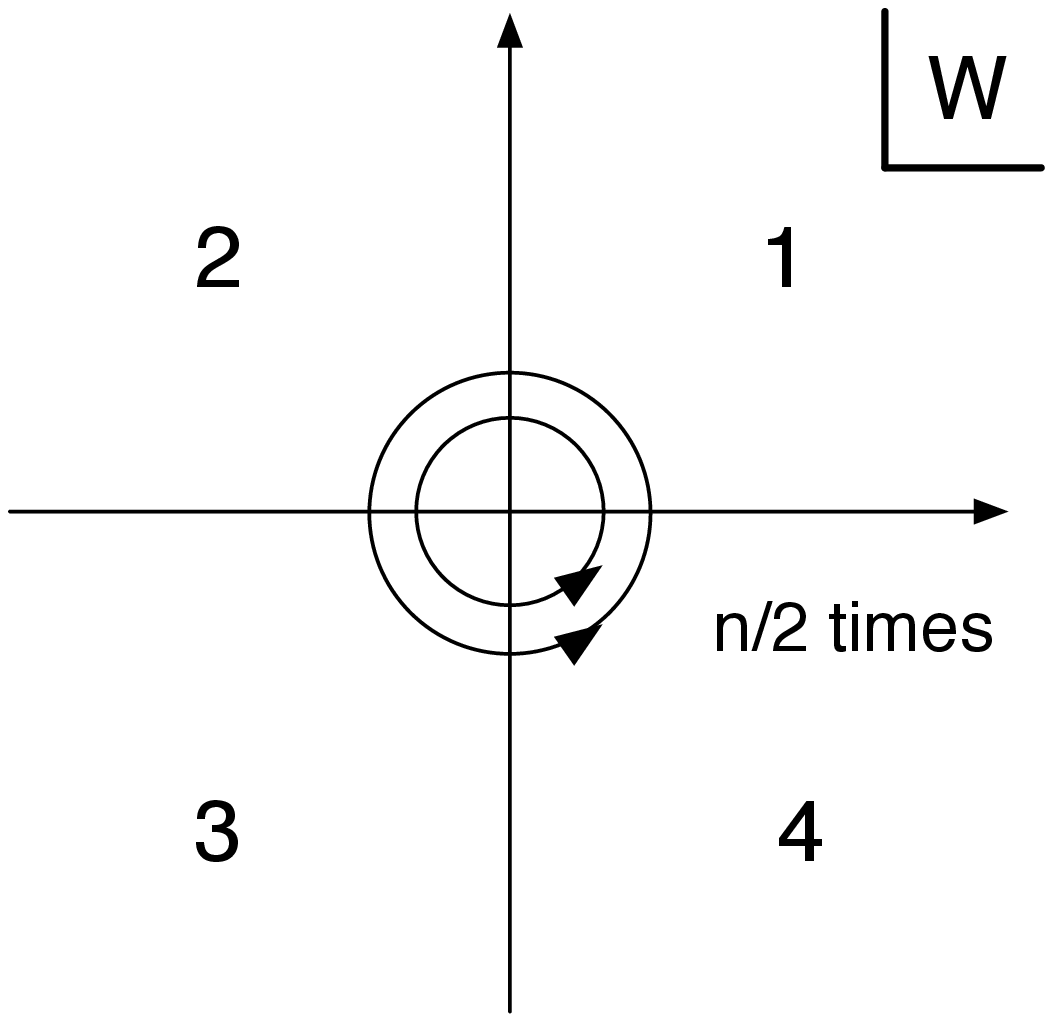}
   \caption{Rotation far from the origin in the $z$- and $w$-planes.}
   \ec
   \vs{-2}
\end{figure}

\subsection{Cross-ratios and area}
The above argument does not say anything about the explicit
form of the solution, and  it is in fact  impossible to obtain
it. Remarkably, it is however possible to extract physical
information without the explicit form of the solution.
Let us see how this is possible.

First, we consider the cross-ratios of the cusp coordinates $x^\mu_i$
related to external momenta. As we go around the $z$-plane,
we pass through regions in the $w$-plane with 
Re$({w}/{\zeta} + \bar{w} \zeta) > 0$ and 
Re$({w}/{\zeta} + \bar{w} \zeta) < 0$ alternatively (Stokes 
sectors). In each region, the linear problem  
has a diverging and a decaying solution as $|w| \to \infty$. 
Let us call them the big and the small solution and denote them
by $b_i$ and $s_i$, respectively. The subscript $i$
labels the region. Explicitly, one has
$b_i, s_i \sim (e^{w/\zeta + \bar{w} \zeta},0 )^t,  
(0,e^{-(w/\zeta + \bar{w} \zeta)})^t$ for large $|w|$, and the solution of
the linear problem is given by 
\eqb
  \psi(\zeta;z) \sim  b_i(\zeta;z) + s_i(\zeta;z) \period \nn
\eqe
It turns out that the cross-ratios are 
expressed by these small solutions as 
\eqb\label{crossratio}
  \frac{ x_{ij}^\pm x_{kl}^\pm }{ x_{ik}^\pm x_{jl}^\pm } 
 = \frac{ (s_i \wedge s_j) (s_k\wedge s_l) }{ (s_i \wedge s_k) (s_j\wedge s_l) }(\zeta)
  =: \chi_{ijkl}(\zeta)
   \comma 
\eqe
where $x_{ij}^{\pm} := x_{i}^{\pm} -x_{j}^{\pm}$, 
$ s_i \wedge s_j :=  \det (s_i, s_j) $, and
 $\zeta = 1$ for $+$ and $\zeta = i$ for $-$.
 Note that $s_i \wedge s_j$
are independent of $z$. This formula relates the geometrical
data of the minimal surfaces carried by $\psi$ to the physical
cross-ratios.

The problem is now how to compute the right-hand side of
(\ref{crossratio}). 
A streamlined solution to this problem is given in \cite{Alday:2010vh}.
There, one first defines the T- and Y-functions by
$$ 
   T_{2k+1} = (s_{-k-1}\wedge s_{k+1}) \comma
    \quad
    T_{2k}  =   (s_{-k-1}\wedge s_{k})^+ \comma 
$$
\vs{-4.5}
$$   
     Y_{s}  = T_{s-1}T_{s+1} \comma  \vs{0.7} 
$$ 
where the superscripts $\pm$ stand for the shift of the argument,
$f^\pm(\zeta) := f(e^{\pm i\pi/2} \zeta)$. Essentially, $Y_s$
are the cross-ratios. For example, $Y_{2k} = -\chi_{-k,k,-k-1,k+1}$. 
By definition, the products $s_i \wedge s_j$ satisfy
the algebraic identity among determinants,
\eqb
 (s_i \wedge s_j) (s_k \wedge s_l) 
  = (s_i \wedge s_k)(s_j \wedge s_l) + (s_i \wedge s_l)(s_k \wedge s_j)
  \period \nn
\eqe
This gives the functional equations among $T_s$ $(s=1, ..., n-3)$,
\eqb\label{Tsystem}
  T_{s}^+T_{s}^- = T_{s+1}T_{s-1}+1 \comma 
\eqe
or in terms of $Y_s$,
\eqb\label{Ysystem}
  Y_s^+ Y_s^- = (1+Y_{s-1})(1+Y_{s+1}) \period 
\eqe
These take the well known form of 
the functional equations which are
called   the T-system/Hirota equations
and the Y-system, respectively.
For a review on T- and Y-systems, see for example \cite{Kuniba:2010ir}.

Up to here,  (\ref{Tsystem}) or (\ref{Ysystem}) is 
just algebraic identities. 
The physical input then comes from the asymptotic behaviors of 
the Y-functions. A WKB analysis of the linear system
(\ref{dpsi}) shows, for example, that
\eqb
  \log Y_{2k} \sim \frac{Z_{2k}}{\zeta} + \log \mu_{2k} \quad (\zeta \to 0) \period
  \nn
\eqe
Here, $Z_s$ are period integrals 
$ Z_s = \oint_{\gamma_s} \sqrt{p} \, dz $, and $\log \mu_s$ are certain
constants, which we call the chemical potentials. In our case of
$AdS_3$, we have $\mu_s =0$. By using such asymptotic behaviors 
and assuming certain analyticity of $ \log Y_s$, one can
convert the Y-system into the following integral equations:
\eqb\label{TBA}
 \log Y_s(\theta) = -m_s R \cosh \theta + K \ast \log(1+Y_{s-1})(1+Y_{s+1})
  \comma 
\eqe
where we have introduced
$\theta := \log \zeta$,  $m_s R := 2 Z_s$, $K(\theta):= 1/\cosh \theta$,
and $\ast$ stands for the convolution, i.e., 
$f \ast g = \int \frac{d\theta'}{2\pi} f(\theta-\theta') g(\theta')$.
For simplicity, we have displayed the equations when all $Z_s$ are real.
The cross-ratios are obtained by solving these equations
and setting the spectral parameter to particular values $\zeta = 1, \pm i$.

The set of equations (\ref{TBA}) are of the form of  
the thermodynamic Bethe ansatz equations, which are used
to analyze finite-size effects of  (1+1) dimensional
integrable systems with factorizable scattering. In 
this context, 
the TBA equations are obtained by minimizing the free energy, 
$m_s$  are the masses  of particles,  $\theta$ is the rapidity, 
$R$ is the inverse temperature, and $\log Y_s$ give pseudo energies.
It is surprising that the geometrical problem of the minimal surfaces
results in the equations of the type of the TBA equations.

The solution to the integral equations (\ref{TBA}) also 
gives the area of the minimal surfaces and hence the
gluon scattering amplitudes at strong coupling.
To see this, we first recall that the area is divergent
and needs to be regularized. Here, we adopt the following
regularization,
\eqb
   A\mbox{\rm (area)} =  4\int d^2z \, e^{2\alpha} \Eqn{\to}
  4\int d^2z \, ( e^{2\alpha} -\sqrt{p\bar{p}} ) + 4 \int_{r \geq \ep} d^2z \, 
   \sqrt{p\bar{p}} \nn \\
  \Eqn{=:} A_{\rm fin} + A_{w\mbox{\scriptsize -vol}} \period \nn
\eqe
With the help of the analysis of the Hitchin system for $\calN=2$ 
SYM, one can find that  the first finite term of the regularized 
area, $A_{\rm fin}$, is nothing but the free energy $F $ associated with
the TBA equations (\ref{TBA}) (up to a sign and a constant):
\eqb\label{Afree}
  A_{\rm fin} \Eqn{=} \sum_s  \int\frac{d\theta}{2\pi}  m_s R \cosh \theta
  \cdot \log(1+Y_s) \ +  \mbox{\rm (const.)} \nn \\
  \Eqn{=} -F +  \mbox{\rm (const.)}  \period
\eqe
On the other hand, the second term, $A_{w\mbox{\scriptsize -vol}} $,
essentially gives the BDS form, 
\eqb
  A_{w\mbox{\scriptsize -vol}} \sim  A_{\rm div} + A_{\rm BDS} + \cdots 
  \comma \nn
\eqe 
where $A_{\rm  div}$ is the divergent part and $A_{\rm BDS}$ is the
finite part of the BDS formula.
Therefore, 
the most intricate part of the remainder function $ {\cal R}$
is given by the free energy: 
\eqb\label{defR}
   {\cal R}  =  -( A- A_{\rm div} - A_{\rm BDS} )  =  
   F + \cdots \period 
\eqe
The ellipses stand for some other terms.

\par\medskip
Summarizing, the procedure of computing the strong-coupling
amplitudes is as follows:

\par\smallskip\noindent
(1)  \ Solve the integral  equations (\ref{TBA}) and obtain the Y-functions
   $Y_s(\theta)$.
   
\par\smallskip\noindent
(2) \  The area $A$ of the minimal surfaces or the amplitude $\calM$ is given
          by the free energy   
           \hspace*{4ex}
            $F$ associated with the TBA equations  and some other terms.

\par\smallskip\noindent     
(3)  \ The cross-ratios 
   (\ref{crossratio}) are obtained
   by evaluating $Y_s(\theta)$ at particular values of the 
   \hspace*{4ex}
     argument
   $\zeta=e^\theta = 1, \pm i$. 

\par\smallskip \noindent    
(4) \ Expressing  the amplitudes by the resultant cross-ratios  gives 
      the amplitudes as   
    \hspace*{4ex}
     functions of  external momenta. 
    
\subsection{Minimal surfaces in $AdS_{5}$}
A similar but more elaborated discussion shows that the minimal surfaces
in $AdS_{5}$ which have $\hat{n}$-cusps are described by   
the following Y-system:   
\eqb\label{AdS5Ysys}
 {Y_{2,m}^-Y_{2,m}^+\over
Y_{1,m}Y_{3,m}}&=&{(1+Y_{2,m+1})(1+Y_{2,m-1})\over(1+Y_{1,m})(1+Y_{3,m})}
\comma \nn
\\
 {Y_{3 ,m}^-Y_{1 ,m}^+ \over
Y_{2,m}}&=&{(1+Y_{3,m+1})(1+Y_{1,m-1})\over
1+Y_{2,m}}  \comma 
\\ 
{Y_{{ 1},m}^-Y_{{ 3}, m}^+ \over
Y_{2,m}}&=&{(1+Y_{1,m+1})(1+Y_{3,m-1})\over 1+Y_{2,m}} \comma
 \nn
\eqe
where  $m=1, .., \hat{n}-5$, and $f^{\pm}(\zeta) = f(e^{\pm i\pi/4} \zeta)$ 
in this case.
This Y-system is non-standard in that  $Y_{1,m}$ and $Y_{3,m}$
couple to each other on the  left-hand side. 
A similar Y-system also appears in the study
of the spectral problem of the AdS${}_{4}$/CFT${}_{3}$ correspondence.

\section{Underlying integrable models and CFTs}

We saw that the minimal surfaces in AdS spaces are
described by the integral equations of the form of the TBA equations
(or the associated Y-systems). A natural question here is:  Are these ``TBA-like''
equations really the TBA equations of  any integrable
models? In the case of the 6-cusp solution in $AdS_{5}$, it has been shown that
the integral equations are indeed the TBA equations of the
$\bbZ_{4}$-symmetric (or $A_{3}$-)integrable model, which is obtained 
by a massive deformation of the $\bbZ_{4}$-parafermion CFT \cite{Alday:2009dv}.   
In the following, we would like to show that the answer to the above question
is yes for the general cusp solutions in $AdS_{3}$ and $AdS_{4}$ \cite{Hatsuda:2010cc}.

Let us first consider the $AdS_{3}$ case. To investigate the 
underlying integrable models, we recall that, if a (1+1)-dimensional 
 integrable model is obtained from a CFT by a relevant perturbation,
the free energy described by the TBA equations  gives
the central charge $c$ of the CFT in the CFT/high-temperature limit
$R \to 0$:
\eqb
   F \to - \frac{\pi}{6} c \period \nn
\eqe
On the other hand, in the same limit, the period integrals $Z_{s}$ are 
vanishing and the minimal surfaces reduce to the regular polygonal
surfaces whose boundary forms a regular polygon in a 
subspace of the AdS boundary after a projection.
This class of the solutions in $AdS_{3}$
are described by the Painlev{\'e} III equation. 
For the $2n$-cusp solution,
the finite  
part of the regularized area in this limit has  been obtained 
as \cite{Alday:2009yn}
\eqb
 A_{\rm fin} \to  \frac{\pi}{4n}(3n^2-8n +4) \period \nn
\eqe
To find the free energy, one has to fix the difference between
$A_{\rm fin}$ and $F$ in (\ref{Afree}). This is done by considering 
another limit where the zeros of the polynomial $p(z)$ become far 
apart from each other. 
Since the solution is expected to be 
a superposition of the $(n-2)$ 6-cusp solutions in this limit, 
it follows that $F \sim 0$ and  $A_{\rm fin} \sim (n-2) \times \frac{7}{12} \pi$.
Thus, 
\eqb
  -F = A_{\rm fin} -\frac{7}{12}(n-2)\pi 
  \to \frac{\pi}{6n}(n-2)(n-3) \period \nn
\eqe

A candidate of the CFT in the  UV/high-temperature limit then has to
have central charge $c=(n-2)(n-3)/n$. One can indeed find such 
a CFT:  The coset or the generalized parafermion CFT associated
with 
\eqb\label{coset}
   \frac{\widehat{\alg{su}}(K)_{k}}{[\widehat{\alg{u}}(1)]^{K-1}} 
\simeq \frac{[\widehat{\alg{su}}(k)_{1}]^{K}}{\widehat{\alg{su}}(k)_{K}}
\eqe
has the central charge
$
  c = {(k-1)K(K-1)}/{(k+K)} \period 
$  
Thus, the coset CFT with $K=n-2$, $k=2$ has the correct central charge.
In addition, the second representation in (\ref{coset}) shows that 
this is an $\alg{su}(2)$ coset, and matches the symmetry of the
$\alg{su}(2)$ Hitchin system. 
Moreover, the degrees of freedom of this coset is $n-3$,
which also matches the number of  independent cross-ratios $2(n-3)$.
We remark that the left and the right sector are described by the same 
integral equations in the $AdS_{3}$ case.

These arguments suggest that the above coset CFT is the right
candidate. Proceeding to a consideration
away from the CFT point,  we note that a massive
deformation of this CFT by the adjoint operators is integrable,
and gives 
 the homogeneous sine-Gordon model \cite{FernandezPousa:1996hi}.
 The model has a factorizable diagonal S-matrix.
In the case of the coset $ \widehat{\alg{su}}(n-2)_{2}/[\widehat{\alg{u}}(1)]^{n-3}$,
the elements of the S-matrix 
 for particles $a$ and $b$ $(a,b = 1, ..., n-3)$ 
are given up to constant factors by
\eqb
  S_{ab}(\theta)  \sim
   \Bigl[  \tanh \half (\theta + \sigma_{ab}- i \frac{\pi}{2})\Bigr]^{I_{ab}}
   \comma \nn
\eqe  
where  $\theta$ is the difference of the rapidities  of the particles, $I_{ab}$ is the 
incidence matrix of $\alg{su}(n-2)$, and  $\sigma_{ab}$ are certain 
parameters.  By the standard procedure, one can then derive the 
TBA equations of this HSG model, to find that they coincide
with the 
integral equations for the minimal surfaces in $AdS_{3}$.
This answers  the question at the beginning of this section affirmatively.
Precisely speaking, the reality of the parameters $\sigma_{ab}$
are different and the physical interpretation should be considered
further. Keeping this in mind, we have found that 
the $2n$-cusp solution
in $AdS_{3}$ is described by the HSG model associated with
the coset  $ {\widehat{\alg{su}}(n-2)_{2}}/{[\widehat{\alg{u}}(1)]^{n-3}}$.
Schematically, 

\vs{1}
\bc
\fbox{\parbox{4.9cm}{ \ $2n$-cusp minimal surfaces \ 
   \vs{0.5}   \\ 
    \hs{10} in $ AdS_3 $   }} 
    \hs{3} $\Longleftarrow$ \hs{3}
    \fbox{\parbox{5.3cm}{ \ \hs{5} HSG model from \ 
   \vs{1}   \\ 
       \hs{2.5} $\displaystyle 
    \frac{\widehat{\alg{su}}(n-2)_{2}}{[\widehat{\alg{u}}(1)]^{n-3}}
     \simeq 
     \frac{[\widehat{\alg{su}}(2)_{1}]^{n-2}}{\widehat{\alg{su}}(2)_{n-2}} $ }} 
\ec
\vs{1}

In the case of $AdS_{5}$, we already know that the 6-cusp 
solution is described by the $\bbZ_{4}$-symmetric integrable model,
which corresponds to the coset (\ref{coset}) with $K=2$ and $k=4$.
Taking into account, again,  the symmetry and the degrees of freedom,
one may 
guess that the $\hat{n}$-cusp solution in $AdS_{5}$ is
described by the HSG model associated with the coset (\ref{coset})
with $K=\hat{n}-4$ and $k=4$. It turns out, however,  
that the TBA equations/Y-system
of this HSG model are of the standard form and do not agree  
with (\ref{AdS5Ysys}). Instead, they do agree with   
those for the $\hat{n}$-cusp solution in $AdS_{4}$,
which are obtained from the $AdS_{5} $ case by setting the chemical potentials 
to  zero and hence identifying $Y_{1,m}$ and $Y_{3,m}$ 
in (\ref{AdS5Ysys}) \cite{Alday:2010vh}. 
The reduction from $AdS_{5}$ to $AdS_{4}$ 
maintains the $\alg{su}(4)$ symmetry of the Hitchin system, and the
identification among the Y-functions ensures the matching between
 the degrees of freedom of the coset and the  number
of independent cross-ratios $2(\hat{n}-5)$.
In addition, the coset CFT has the central charge 
$c=3(\hat{n}-4)(\hat{n}-5)/\hat{n}$,
which also agrees with the result for the regular polygon solution
in $AdS_{4}$ \cite{Alday:2010vh}. Schematically,

\vs{1}
\bc
\fbox{\parbox{4.9cm}{ \ \hs{1} $\hat{n}$-cusp minimal surfaces \ 
   \vs{0.5}   \\ 
    \hs{10} in $ AdS_4 $   }} 
    \hs{3} $\Longleftarrow$ \hs{3}
    \fbox{\parbox{5.4cm}{ \ \hs{5}  HSG model from  \ 
   \vs{1}   \\ 
       \hs{2.5} $\displaystyle 
    \frac{\widehat{\alg{su}}(\hat{n}-4)_{4}}{[\widehat{\alg{u}}(1)]^{\hat{n}-5}}
     \simeq 
   \frac{[\widehat{\alg{su}}(4)_{1}]^{\hat{n}-4}}{\widehat{\alg{su}}(4)_{\hat{n}-4}} $ }} 
\ec
\vs{1}

The reduction from $AdS_{5}$  to $AdS_{4}$  
seems to suggest a possibility that their Y-systems are related by 
certain deformations of the underlying CFT/integrable model
by the chemical potentials.
We see a simple example of such a deformation 
in the case of the 6-point amplitudes. 
The identification of the underlying 
CFT and integrable model in the $AdS_{5}$ case is an interesting issue
to be discussed further. As a side remark, we note that
the $\bbZ_{4}$-symmetric integrable model for the 6-cusp solution
is a special case of the HSG model.

\section{Six-point  amplitudes from $Z_4$-symmetric integrable model}

We saw that the HSG model associated with certain cosets
describes the minimal surfaces in AdS spaces and hence
the scattering amplitudes at strong coupling. This implies 
an unexpected  connection between a four-dimensional
SYM theory and (1+1)-dimensional integrable models. 
Such an identification  
is  not only interesting but also useful in analyzing the amplitudes.
We would like to demonstrate this 
in the case of the 6-point amplitudes corresponding to 
 the 6-cusp solution in $AdS_{5}$ by deriving their expansion 
near the CFT limit  
\cite{Hatsuda:2010vr}.

As mentioned before, 
the 6-cusp minimal surfaces in $AdS_{5}$ are described by the
TBA equations associated with the $\bbZ_{4}$-symmetric  integrable model.
The model is obtained by an integrable deformation of the 
$\bbZ_{4}$-parafermion CFT by the first energy operator $\varep(x)$
with dimension $D_{\varep} = \bar{D}_{\varep} = 1/3$.
Its action is given by
\eqb
  S = S_{PF} + g \int d^2x \, \varep(x) \comma \nn
\eqe  
where $S_{PF}$ is the action of the $\bbZ_{4}$-parafermion CFT, 
which has the central charge $c=1$. The model contains three particles
with mass $ m_{a} =m, \sqrt{2} m$ and $m$, respectively. The  third
particle is the anti-particle of the first. The coupling constant 
$g$ is related to the  mass as $g = b_{g} m^{4/3}$
with $b_{g}$ being a  certain numerical constant. 
The Y-system of this model is
\eqb\label{Z4Ysys}
   Y^{+}_{1} Y^{-}_{1} =  1+ Y_{2} \comma \quad 
  Y^{+}_{2} Y^{-}_{2} = (1+ \mu Y_{1}) (1+\mu^{-1}Y_{1}) 
   \comma  \quad Y_{1} = Y_{3} \comma
\eqe
where $f^{\pm}(\theta) = f(\theta \pm \frac{\pi}{4}i)$ and 
$\log \mu$ is the chemical potential.
As in the $AdS_3$ case, this can be converted to
the TBA equations.
In the following, we discuss the amplitudes around the CFT limit
where all the independent cross-ratios 
are equal in a certain basis.

First, let us consider the free energy associated with the TBA equations.
In general, the free energy of a model on a circle of length $L \gg 1$ with 
temperature $1/R$ gives  the 
ground state energy $E(R)$ of the model on a circle of length $R$.
This relation is found by evaluating the torus partition function in two different 
channels. 
Near the CFT/high-temperature limit $mR := 2 |Z| \ll 1$, 
the CFT perturbation gives an expansion of the free energy.
In our case, it reads
\eqb\label{Fexpansion}
   \qquad F \Eqn{=}  E_0 + \frac{1}{4}(mR)^2 - R^2 
   \sum_{n=1}^{\infty} \frac{(-g)^{n}}{n!} 
     \Bigl( \frac{2\pi}{R}\Bigr)^{2(D_{\varep}-1)n +2}   \\
  && \times  \int \Big\bra V(\infty) \, \varep(z_{n},\zbar_{n}) 
   \cdots \varep(z_{1},\zbar_{1}) \, V(0) \Big\ket_{\rm CFT} 
   \prod_{i=2}^{n} (z_{i}\zbar_{i})^{D_{\varep}-1} dz^{2}_{2} \cdots dz^{2}_{n}
   \comma \nn
\eqe
where $E_{0}$ is the CFT ground state energy $-\pi/6$ and $V$ is the vacuum
operator. The correlators are connected ones of the CFT  on a complex plane, 
and we have set $z_{1} =1$. $|Z|$ is the absolute value of a
period integral $ Z =: |Z| e^{i\varphi}$ 
similar to $Z_{s}$ in the $AdS_{3}$ case.

When the chemical potential vanishes, i.e., $\mu =: e^{i\phi}=1$, the vacuum
operator is the identity, $V= 1$. On the other hand,  $\mu \neq 1$
corresponds to a twisted boundary condition of the 
$\bbZ_{4}$-parafermion CFT. The vacuum operator in this case 
becomes non-trivial. 
To be explicit, we bosonize the parafermion 
theory by a free boson $\Phi$. The energy and the vacuum operator
are then given by
\eqb
   \varep = a_+ \, e^{i \sqrt{\frac{2}{3}} \Phi} +  a_-e^{-i \sqrt{\frac{2}{3}} \Phi}
   \comma \quad 
   V = e^{-i\sqrt{\frac{1}{6}}\frac{\phi}{\pi}\Phi} \comma \nn
\eqe
where $a_{\pm}$ are certain cocycle factors. Substituting these into the
expansion (\ref{Fexpansion}), one obtains
\eqb\label{6ptFexpansion}
  F =  E_0 + |Z|^2 - C_{\frac{8}{3}} \,
  \gamma\Bigl(\frac{1}{3} +\frac{\phi}{3\pi}\Bigr) 
    \gamma\Bigl(\frac{1}{3} -\frac{\phi}{3\pi}\Bigr) \, |Z|^{\frac{8}{3}}
    + \calO ( |Z|^{\frac{16}{3} } ) \comma 
\eqe
where 
$E_0 = -\frac{\pi}{6} (1- \frac{2\phi^2}{\pi^2})$, $\gamma(z) 
= \Gamma(x)/\Gamma(1-x)$,
and  $C_{\frac{8}{3}}= \frac{\pi}{2} 
   \bigl[\frac{1}{\sqrt{\pi}} \gamma\bigl(\frac{3}{4}\bigr)\bigr]^{\frac{8}{3}}
      \gamma\bigl(\frac{1}{6}\bigr) \gamma\bigl(\frac{1}{3}\bigr)  
      \approx 0.18461 $.
      We have also used the explicit value of $b_{g}$.
This is in good agreement with numerical computations (fig. 4).
\begin{figure}[t]
   \bc
   \includegraphics*[width=7.3cm]{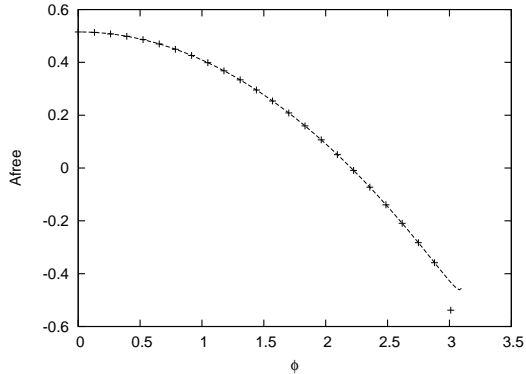}
   \caption{Plot of $A_{\rm free} :=-F $ as a function of 
   $\phi$  
    for $|Z| = 0.1$ and $\varphi=-\pi/48$  from 
     numerical results ($+$) and 
     the high-temperature expansion (\ref{6ptFexpansion}) (dashed line).}
     \ec
     \vs{-2}
\end{figure}

Besides the free energy, one can also find an expansion of the
Y-functions. 
From the periodicity 
and the analyticity, the Y-functions are expanded as
\eqb
  Y_a(\theta) =  \sum_{n=0}^{\infty}  Y_a^{(n)} \cosh\Bigl( \frac{4}{3} n(\theta -i\varphi) \Bigr)
  \comma \nn
\eqe
where   $Y_a^{(n)} \sim (mR)^{4n/3} $ as $mR \to 0$.
Substituting the expansion into the Y-system (\ref{Z4Ysys})
gives equations to constrain  the coefficients $Y_{a}^{(n)}$. 
Further using the relation between the Y-functions and the cross-ratios,
$
   U_{k} = 1+ Y_{2} \bigl(\frac{2k+1}{4} \pi i  \bigr) $ $ (k=1,2,3) $,
one finds the first-order expansion around equal $U_{k}$:
\eqb
  U_k=4 \cos^{2}\bigl(\frac{\phi}{3} \bigl)
+y^{(1)}(\phi) \cos\Bigr(\frac{4\varphi-(2k+1)\pi}{3} \Bigr) \times 
|Z|^{\frac{4}{3}}
+{\cal O}\bigl(|Z|^{\frac{8}{3}}\bigr) \comma \nn
\eqe
where
$y^{(1)}$ is a function of the chemical potential $\phi$.  Numerically, 
this is evaluated 
as  $ y^{(1)}(\phi) \approx5.47669  -0.484171\phi^2  + 0.0119471\phi^4 
+ \cdots $.
The above relations are inverted 
to express the data of the minimal surfaces 
($|Z|, \varphi, \phi$) 
as functions of the cross-ratios,
\eqb\label{UkZphi}
    \cos^2 \frac{\phi}{3} = \frac{1}{12} \sum_{k} U_{k}
    \comma \quad 
   \tan \frac{4}{3} \varphi  =
   \frac{\sqrt{3}(U_2-U_3)}{2U_1 -U_2 -U_3} \comma 
   \quad
    |Z|^{\frac{4}{3}}  = 
    \frac{-2U_1 +U_2 +U_3}{3 y^{(1)}(\phi) \cos \frac{4}{3} \varphi} \period  
\eqe
From these,   geometrical meaning 
of $ (|Z|,\varphi,\phi)$  in the parameter space $(U_1,U_2,$ $U_3)$
is  found.

Collecting all the results in addition to (\ref{6ptFexpansion}), the full expression 
of the remainder function defined in (\ref{defR})
is found to be
\eqb\label{Rexpansion}
{\cal R} \! \! \Eqn{=} \! \! \!
 -\left[\frac{\pi}{6}\Bigl(1-\frac{2\phi^2}{\pi^2}\Bigr)
  +\frac{3}{4}\dilog(1-4\beta^2)\right]  \\
&& \! \! \!  -\left[ 
  C_{\frac{8}{3}}\gamma\Bigl(\frac{1}{3}+\frac{\phi}{3\pi}\Bigr)
  \gamma\Bigl(\frac{1}{3}-\frac{\phi}{3\pi}\Bigr)
  -\frac{3\bigl(4\beta^2-1+\log(4\beta^2)\bigr)}{64\beta^2(4\beta^2-1)^2}
  y^{(1)}(\phi)^2
  \right]|Z|^{\frac{8}{3}}
  +{\cal O}\bigl(|Z|^4\bigr) \comma \nn 
\eqe
where $\dilog$ is the dilogarithm and $\beta:= \cos(\phi/3)$.
By  (\ref{UkZphi}), this is further expressed
in terms of the cross-ratios $U_{k}$, which can be directly compared
with perturbative computations.

In addition to the above expansion around the CFT limit with $|Z| \ll 1$,
it is straightforward to carry out the opposite expansion
around the low-temperature/infrared limit with $|Z| \gg1$, which corresponds
to collinear limits in the SYM theory.
In fig. 5, we show  the remainder function obtained 
by the first order expansions for $|Z| \ll 1$ and $|Z| \gg1$.
These are again in good agreement with numerical computations.
We find that the simple first order expansions
well describe the remainder function for all the scale $|Z|$.

\begin{figure}[t]
   \bc
   \includegraphics*[width=7.6cm]{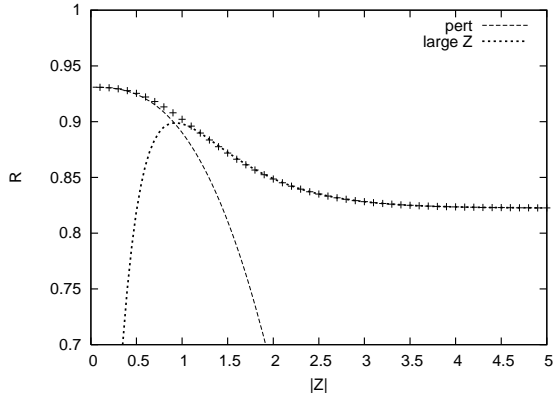}
   \caption{Plot of the remainder function $\cal R$ as a function of $|Z|$ 
   for $\phi = 0$ and $\varphi= -\pi/48$
   from the UV expansion (\ref{Rexpansion}) (dashed line), 
   the first order IR expansion (dotted line) and numerical results ($+$).}
    \ec
    \vs{-2}
\end{figure}

\section{Summary}

The discovery of the integrability 
opened up new dimensions 
in the study of  the gauge/ string duality or the AdS/CFT correspondence.
That has led to a proposal that the full spectrum of $\calN =4$ SYM
and the string theory on $AdS_{5} \times S^{5}$ in the planar limit 
is obtained by solving certain TBA equations/Y-system \cite{Tateo-san}.
Besides this very interesting theoretical development,
the integrability has also been applied to
the study of the gluon scattering amplitudes:
By the AdS/CFT correspondence,
the amplitudes at strong coupling are given by the area of the minimal surfaces
in  $AdS_{5}$ 
with a polygonal boundary which consists of null edges
corresponding to external momenta \cite{Alday:2007hr}. 
These minimal surfaces are again described 
by certain, but different, TBA equations/Y-system 
\cite{Alday:2009yn,Alday:2009dv,Alday:2010vh,Hatsuda:2010cc}. 
The study on the strong coupling side also provided useful insights
and facilitated the development  on the weak coupling side.

Generalizing the connection between the 6-point amplitudes
and the $\bbZ_{4}$-symmetric integrable model \cite{Alday:2009dv},
we observed that the TBA equations/Y-systems for the minimal
surfaces in $AdS_{3}$ and $AdS_{4}$ coincide with 
those of the HSG model associated with certain coset or generalized
parafermion CFTs \cite{Hatsuda:2010cc}. Such a connection between 
the scattering 
amplitudes of the four-dimensional SYM theory and the (1+1)-dimensional
integrable model is not only interesting but also 
useful for actual computation of the amplitudes. We demonstrated
this in the case of the 6-point amplitudes by deriving 
an expansion near the limit of equal cross-ratios
from the $\bbZ_{4}$-symmetric integrable model \cite{Hatsuda:2010vr}.

There may be many future directions.
First, although we arrived at the TBA equations
/Y-systems in analyzing
the amplitudes, the intrinsic reason is not clear. The situation 
resembles that of the ODE/IM correspondence \cite{Dorey:2007zx}. 
Second, the integrable model underlying the $AdS_{5}$ case
is not yet identified, except for the 6-point case. 
It would be interesting to clarify whether
it is just  some deformation of the $AdS_{4}$ case by 
the chemical potentials as in the 6-point case, or 
corresponds to some new integrable model. Third, it seems that we do not
have a formalism to  obtain 
the completely analytic form of  the expansion of the Y-functions 
near the CFT limit. This is important to derive an analytic 
form of the remainder function, though in some restricted parameter space.
Given the recent development on the analytic form of the perturbative
remainder function, this is certainly an interesting issue.
Finally, recalling the development on the spectral problem
where the expansions both from the weak and strong coupling 
side finally reached the proposal of the full spectrum,
it would be very interesting if one could include the 
corrections to the strong coupling result. (See the last figure.)
On the day before this talk was given, an interesting  paper  
\cite{Alday:2010ku} appeared
which discusses this issue.

\begin{figure}[t]
   \bc
   \includegraphics*[width=12cm]{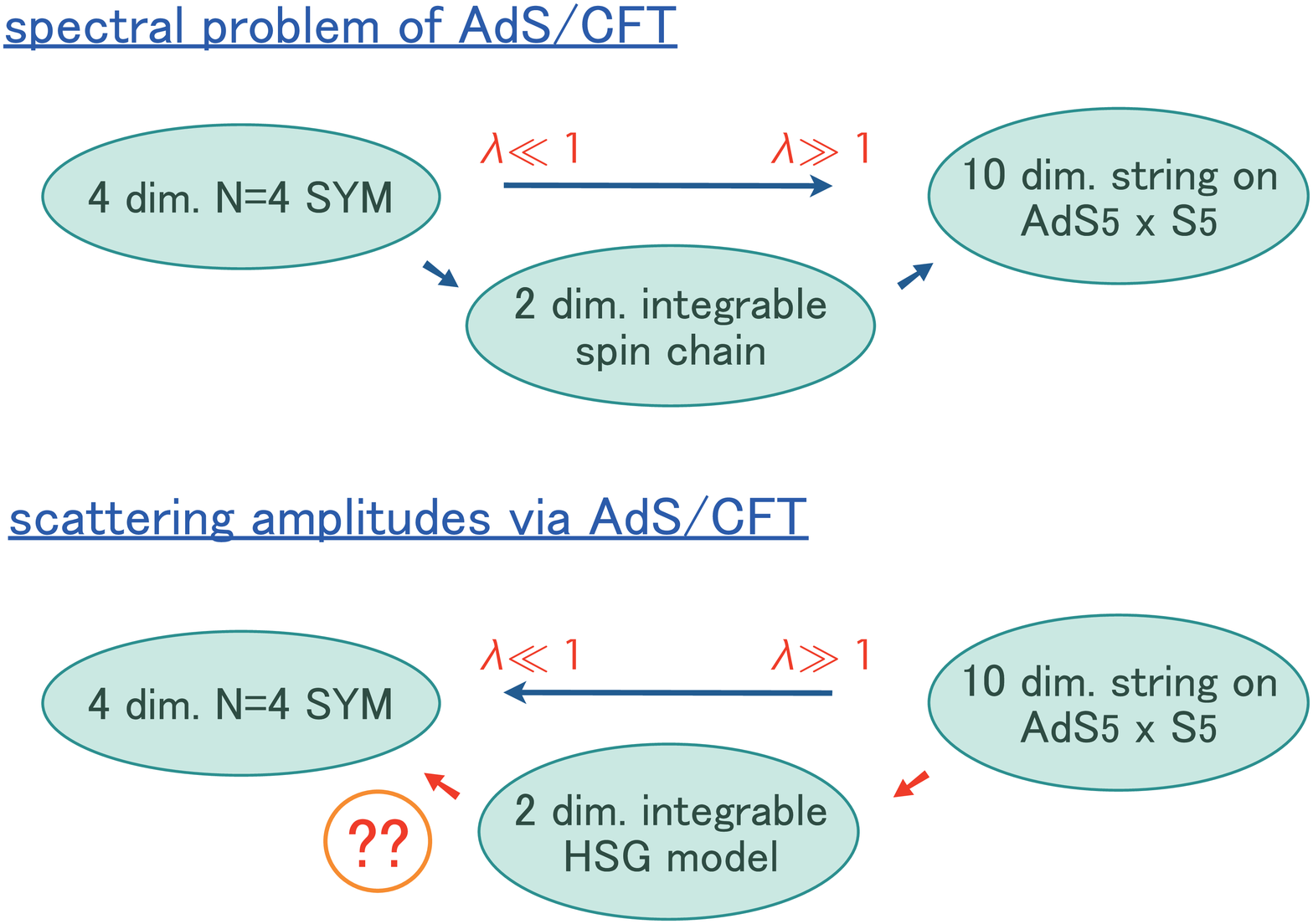}
   \ec
   \vs{-4}
\end{figure}

 \section{Acknowledgements}
 The author would like to thank the organizers 
 for giving him an opportunity to speak in this stimulating workshop.
 He would also like to thank Y. Hatsuda, K. Ito and K. Sakai
 for fruitful collaborations on this subject, and J. Suzuki 
 and R. Tateo for very useful discussions. 
This work is supported in part by Grant-in-Aid
for Scientific Research from  
Ministry of Education, Culture, Sports, Science and Technology.

%
%
\def\thebibliography#1{\list
 {[\arabic{enumi}]}{\settowidth\labelwidth{[#1]}\leftmargin\labelwidth
  \advance\leftmargin\labelsep
  \usecounter{enumi}}
  \def\newblock{\hskip .11em plus .33em minus .07em}
  \sloppy\clubpenalty4000\widowpenalty4000
  \sfcode`\.=1000\relax}
 \let\endthebibliography=\endlist
%
%
\vspace{3ex}
\begin{center}
 {\large\bf References}
\end{center}
\par 

%

%
\end{document}